\documentclass[superscriptaddress,twocolumn]{revtex4}
\usepackage{amssymb}
\usepackage[tbtags]{amsmath}
\usepackage{graphicx}
\usepackage{epsfig,graphicx,times}

\begin{document}

\title{Testing Bell's inequality in constantly coupled Josephson
circuits by effective single-qubit operations}
\author{L.F. Wei}
\affiliation{Frontier Research System, The Institute of Physical
and Chemical Research (RIKEN), Wako-shi, Saitama, 351-0198, Japan}
\affiliation{Institute of Quantum Optics and Quantum Information,
Department of Physics, Shanghai Jiaotong University, Shanghai
200030, P.R. China}
\author{Yu-xi Liu}
\affiliation{Frontier Research System, The Institute of Physical and Chemical Research
(RIKEN), Wako-shi, Saitama, 351-0198, Japan}
\author{Franco Nori}
\affiliation{Frontier Research System, The Institute of Physical
and Chemical Research (RIKEN), Wako-shi, Saitama, 351-0198, Japan}
\affiliation{Center for Theoretical Physics, Physics Department,
CSCS, The University of Michigan, Ann Arbor, Michigan 48109-1120,
USA}
\date{\today }

\begin{abstract}
In superconducting circuits with interbit untunable (e.g.,
capacitive) couplings, ideal local quantum operations cannot be
exactly performed on individual Josephson qubits. Here we propose
an effective dynamical decoupling approach to overcome the
``fixed-interaction" difficulty for effectively implementing
elemental logical gates for quantum computation. The proposed
single-qubit operations and local measurements should allow
testing Bell's inequality with a pair of capacitively-coupled
Josephson qubits. This provides a powerful approach, besides
spectral-analysis~[Nature \textbf{421}, 823 (2003); Science
\textbf{300}, 1548 (2003)], to verify the existence of macroscopic
quantum entanglement between two fixed-coupling Josephson qubits.

\vspace{0.3cm} PACS number(s): 03.65.Ud, 03.67.Lx, 85.25.Dq.
\end{abstract}

\maketitle

\section{Introduction}
Non-locality (i.e., entanglement)
is one of the most profound features of quantum theory and plays
an important role in quantum information processing, including
quantum computation and quantum communication~\cite{NC00}.
Mathematically, an entangled pure state of a composite system is
defined as a state that cannot be factorized into a direct tensor
product of the states associated with individual subsystems. The
presence of multi-partite entanglement is necessary for
implementing quantum algorithms that are exponentially faster than
classical ones~\cite{NC00}. Also, entanglement may offer some new
information transfer modes, e.g., teleportation and quantum
cryptography~\cite{NC00}. Therefore, generating and verifying
entanglement between qubits are of great practical
importance~\cite{Horodecki02}.

The existence of entanglement can be verified by using various
experimental methods, e.g., quantum tomographic techniques,
Bell-state analysis, quantum jump measurements, etc. (see,
e.g.,~\cite{Roos04,Tsai03-2}). Also, spectral analysis has been
used to probe the existence of two-qubit entanglement in coupled
Josephson qubits~\cite{Tsai03, Berkely03}. However, the degree of
entanglement between the two always-present interacting qubits
changes rapidly~\cite{Abouraddy01} and, at certain times,
two-qubit states can be almost separable. Thus, verifying the
instantaneously generated entangled state in coupled
systems~\cite{W03}, and using it to realize quantum information
processing, e.g., teleportation and quantum memory, are very
important challenges.

Historically, Bell's inequality always served as one of the most
important witnesses of entanglement: its violation implies that
entanglement must be shared by the separate parts. Numerous
experimental tests of Bell's inequality have been made with
entangled photons separated far apart (e.g., up to $500$
m)~\cite{asp82} and entangled {\it closely-spaced} trapped ions
(e.g., {\it separated a few micrometers apart})~\cite{Rowe01}. A
Bell-like inequality has also been tested via single-neutron
interferometry by measuring the correlations between two entangled
degrees of freedom (comprising spatial and spin components) of
{\it single} neutrons~\cite{Hasegawa03}. The results from these
experiments strongly violate the tested Bell's inequalities, and
thus agree with quantum mechanical predictions. Very recently,
preliminary proposals have been explored for testing Bell's
inequalities with switchable Josephson qubits~\cite{Wei04}.

Almost all proposals for manipulating quantum information rely on
the execution of both single-qubit and two-qubit gates. In some
cases (e.g., trapped ions, QED cavities, and quantum
dots~\cite{CZ95}), single-qubit operations are easy to realize by
applying certain controllable local fields. However, the interbit
couplings are fixed and uncontrollable in current experimental
Josephson circuits~\cite{Tsai03,Tsai03-2,Berkely03} and NMR
systems~\cite{GC}. Nevertheless, qubits in NMR are individually
addressable, because the coupling constants $J_{ij}$ between the
$i$th and $j$th qubits are sufficiently weak (i.e., much smaller
than the differences $\Delta\omega_{ij}=|\omega_i-\omega_j|$
between the eigenfrequencies of the qubits, e.g.,
$J_{ij}/\Delta\omega_{ij}\lesssim 10^{-4}$~\cite{GC}). However,
the usual capacitive coupling in Josephson
circuits~\cite{Tsai03-2,Tsai03} is relatively strong, thus making
it difficult to perform local single-qubit operations on
individual qubits. The method~\cite{Lidar02} of physically
arranging qubits (i.e., using two or three physical qubits to
encode a logical qubit) cannot be directly used in the present
two-qubit system.

Here we develop a new approach to overcome the serious
``fixed-interaction" difficulty and effectively implement desired
single-qubit operations on selected qubits. This dynamical
decoupling method can be used to generate long-lived entangled
states in coupled Josephson qubits. The long-lived entanglement
obtained here, assisted by the proposed local single-qubit
operations, should allow to test Bell's inequality with a pair of
capacitively coupled Josephson qubits. Its violation would provide
another robust physical evidence of statistically nonlocal
correlations at the macroscopic scale.

\begin{figure}[tbp]
\vspace{4cm}
\includegraphics[width=13.6cm]{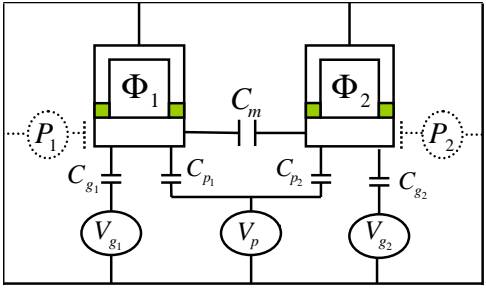}
\vspace{-8cm} \caption{Two capacitively coupled charge qubits. The
quantum states of two Cooper-pair boxes (i.e., qubits) are
manipulated by controlling the applied gate voltages
$V_{g_1},\,V_{g_2}$ and external magnetic fluxes
$\Phi_{1},\,\Phi_{2}$ (penetrating the SQUID loops). $P_{1}$ and
$P_{2}$ (dashed line parts) read out the final qubit states.}
\end{figure}

\section{Dynamical decoupling}

We consider the
nano-circuit sketched in Fig. 1. This is similar to that in
Ref.~\cite{Tsai03}, just replacing the left Josephson junction
there by a superconducting quantum interference device (SQUID)
loop with controllable Josephson energy. The two Cooper-pair boxes
are coupled via the capacitance $C_m$. The qubits work in the
charge regime, with $k_BT\ll E^{(j)}_J\ll
E^{(j)}_C\ll\Delta$\,($j=1,2$), where both quasi-particle
tunnelling and
excitations are effectively suppressed and the number $n_j$\,(with $%
n_j=0,1,2,...$) of Cooper-pairs in the boxes is a good quantum
number. Here, $k_B,\,T,\,\Delta,\,E^{(j)}_C$, and $E^{(j)}_J$ are
the Boltzmann constant, temperature, superconducting gap, the
charging and Josephson energies of the $j$th qubit, respectively.
Following Refs.~\cite{Tsai03-2,Tsai03}, the system is assumed to
work near the co-resonant point and its quantum dynamics can be
restricted to the subspace spanned by the four lowest charge
states: $|00\rangle,|10\rangle,|01\rangle$ and $|11\rangle$. Thus,
the Hamiltonian of this circuit is
\begin{eqnarray}
\hat{H}=\frac{1}{2}\sum_{j=1,2}\left[E^{(j)}_C\sigma^{(j)}_{z}-E^{(j)}_J%
\sigma^{(j)}_{x}\right]+E_{12} \; \sigma^{(1)}_{z}\sigma^{(2)}_z.
\end{eqnarray}%
Here, $E_{12}=E_m/4$ is effective interbit-coupling with
$E_m=4e^{2}C_{m}/C_\Sigma$. The effective charge energy
$E^{(j)}_C$ is
$E^{(j)}_C=E_{C_j}\left(-1/2+n_{g_j}\right)+E_{m}\left(-1/4+n_{g_k}/2\right)
\,\,(j,k=1,2)$ with $ n_{g_j}=(C_{g_j}V_{g_j}+C_{p_j}V_{p})/(2e)$
and $ E_{C_j}=4e^{2}C_{\Sigma_{k}}/C_\Sigma$. The effective
Josephson energy of the SQUID is
$E^{(j)}_J=2\varepsilon_J^{(j)}\cos\left(\pi\Phi_j/\Phi_0\right)$
with Josephson energy $\varepsilon^{(j)}_{\rm J}$ of the single
junction. Above,
$C_\Sigma=C_{\Sigma_{1}}C_{\Sigma_{2}}-C_{m}^{2}$, and
$C_{\Sigma_j}$ is the sum of all capacitances connected to the
$j$th box. The pesudospin operators are defined as
$\sigma^{(j)}_z=|0_{j}\rangle \langle 0_{j}|-|1_{j}\rangle \langle
1_{j}|$ and $\sigma^{(j)}_{x}=|0_{j}\rangle \langle
1_{j}|+|1_{j}\rangle \langle 0_{j}|$.

First, let us consider the circuit working at the co-resonance
point (i.e., $n_{g_{1}}=n_{g_{2}}=0.5$, yielding
$E^{(1)}_C=E^{(2)}_C=0$), and the applied fluxes are set as
$\Phi_j$$=$$0,\,\Phi_k$$=$$\Phi_0/2,\,k\neq j$ (yielding
$E^{(j)}_J$$=$$2\varepsilon_J^{(j)},\,E^{(k)}_J$$=$$0$). In this
case, the circuit has the Hamiltonian
\begin{equation}
\hat{H}_1=-\varepsilon^{(j)}_J\sigma^{(j)}_x+E_{12} \;
\sigma_z^{(1)}\sigma_z^{(2)}.
\end{equation}
The corresponding time-evolution operator reads
\begin{equation}
\hat{U}_1(t)=\exp\left(-\frac{it}{\hbar}\hat{H}_1\right)=\exp\left[\frac{it}{\hbar}\varepsilon_J^{(j)}
\sigma^{(j)}_x\right]\hat{U}_{\rm int}(t),
\end{equation}
where the operator $\hat{U}_{\rm int}(t)$ is determined by
\begin{equation}
i\hbar\frac{\partial\hat{U}_{\rm int}(t)}{\partial t}
=\hat{H}_{\rm int}(t)\hat{U}_{\rm int}(t),
\end{equation}
with $$\hat{H}_{\rm
int}(t)=E_{12}\exp\left[-\frac{it}{\hbar}\varepsilon_J^{(j)}\sigma^{(j)}_x\right]\sigma^{(1)}_z\sigma^{(2)}_z
\exp\left[\frac{it}{\hbar}\varepsilon_J^{(j)}\sigma^{(j)}_x\right].$$
Considering $\zeta_j=E_{12}/(2\varepsilon_J^{(j)})\ll 1$ (e.g.,
$\zeta_j\lesssim 1/4$ in the circuit~\cite{Tsai03}), one can make
the following perturbation expansion:
\begin{widetext}
\begin{eqnarray}
\hat{U}_{\rm int}(t)&=&1+\left(-\frac{i}{\hbar}\right)\int^{t} dt^{\prime}\, \hat{%
H}_{\rm int}(t^{\prime})+\left(-\frac{i}{\hbar}\right)^2\int^{t}\int^{t^{\prime}}dt^{\prime}d{%
t^{\prime\prime}} \hat{H}_{\rm int}(t^{\prime})\,\hat{H}_{\rm int}(t^{\prime%
\prime})+...\nonumber\\
&=&1-\left(\frac{it}{\hbar}\right)\times\frac{E^2_{12}}{2\varepsilon_J^{(j)}}\,\sigma^{(j)}_x\otimes
I^{(k)}+\hat{O}(\zeta^2_j).
\end{eqnarray}
\end{widetext}
Neglecting the higher-order terms of $\zeta_j$, since it is small,
the Hamiltonian of the system can be effectively rewritten as
\begin{equation}
\hat{H}_{\rm
eff}^{(j)}=-\left[\varepsilon_J^{(j)}+\frac{E^2_{12}}{2\varepsilon_J^{(j)}}\right]\sigma
_{x}^{(j)}\otimes\hat{I}^{(k)}.
\end{equation}
Above, the first-order expansion term $\hat{U}^{(1)}_{\rm
int}(t)=(-i/\hbar)\int^{t} dt^{\prime} \hat{%
H}_{\rm int}(t^{\prime})$ practically does not contribute to the
time evolution, due to its small probability (proportional to
$\zeta^2_j$). Under this approximation, the fixed interaction
between the qubits has been effectively eliminated, except
resulting in a shift of the relatively strong Josephson energy.
Thus, the system effectively undergoes an evolution:
\begin{equation}
\hat{R}_{x}^{(j)}(\varphi_j)=\exp \left[i\varphi_j\,\sigma _{x}^{(j)}\right]%
,\ \
\varphi_j=\frac{\varepsilon_J^{(j)}t}{\hbar}\left(1+2\zeta_j^{2}\right),
\end{equation}
which reduces to the single-qubit $\sigma _{x}^{(j)}$-rotation
(i.e., qubit-flip) on the $j$th qubit, if the duration is set by
$\cos\varphi_j=0$.

The robustness of this dynamical decoupling can be verified by
testing the difference of the corresponding physical effects,
e.g., the transition probabilities $P$ between two selected
states, due to the present approximate time-evolution
\begin{equation}
\hat{U}_{\rm appr}(t)
=R_x^{(j)}(\varphi_j)\otimes I^{(k)}
\end{equation}
and the exact one
\begin{eqnarray}
\hat{U} _{\rm ex}(t)&=&\exp(-it\hat{H}_1/\hbar)
=A(t)\sigma^{(j)}_x\otimes \hat{I}
^{(k)}\nonumber\\&+&B(t)(|0_j0_k\rangle\langle 0_j0_k|
+|1_j1_k\rangle\langle
1_j1_k|)\nonumber\\&+&B^*(t)(|1_j0_k\rangle\langle
1_j0_k|+|0_j1_k\rangle\langle 0_j1_k|),
\end{eqnarray}
respectively. Here,
\begin{equation}
A(t)=i\rho_j(t),\,B(t)=[1-\rho_j^2(t)]^{1/2}\,\exp[-i\xi_j(t)],
\end{equation}
with
\begin{eqnarray*}
\rho_j(t)&=& \nu_j^{-1}\sin(\varepsilon_J^{(j)}\nu_j t/\hbar), \\
\nu_j   &=& [1+(E_{12}/\varepsilon_J^{(j)})^2]^{1/2}, \\
\xi_j(t)&=&
\arctan[\,2\zeta_j\nu_j^{-1}\tan(\varepsilon_J^{(j)}\nu_j t
/\hbar)].
\end{eqnarray*}
\begin{figure}[tbp]
\vspace{1.3cm}
\includegraphics[width=12.6cm]{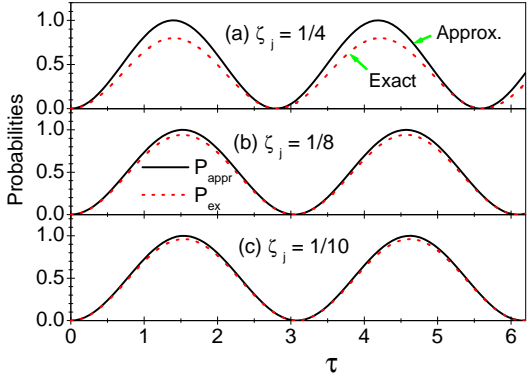}
\vspace{-4.5cm} \caption{Probabilities $P_{\rm appr}$ (solid
lines) and $P_{\rm ex}$ (dotted lines) for the transition
$|1_j0_k\rangle\leftrightarrow |0_j0_k\rangle$ versus
$\tau=\varepsilon^{(j)}_Jt/\hbar$, due to the approximate and
exact time evolutions, respectively. The difference between them
decreases with decreasing interbit coupling $\protect\zeta_j$.}
\end{figure}
Figure 2 shows that the probabilities for the transition
$|1_j0_k\rangle\leftrightarrow |0_j0_k\rangle$ due to the
evolutions $\hat{U}_{\rm appr}(t) $ (solid lines) and
$\hat{U}_{\rm ex}(t)$ (dotted lines) possess the same oscillating
period. Also, the difference between the probabilities decreases
when deceasing the coupling strength $\zeta_j$: the largest
differences are less than $0.06$ and $0.04$ for coupling strengths
$\zeta_j=1/8$ and $\zeta_j=1/10$, respectively.

Similarly, if the system works far from the co-resonance point
(e.g., $n_{g_{j}}<0.25$,~\cite{Tsai03-2}), then both $E_J^{(j)}$
and the fixed coupling $E_{12}$ are the small perturbative
quantities, compared to the charging energy $E_C^{(j)}$. Thus, the
Hamiltonian (1) can be effectively approximated by
\begin{eqnarray}
\hat{H}_{2}=\sum_{j=1,2}E_{j}\,\sigma _{z}^{(j)}+E_{12}\,\sigma
_{z}^{(1)}\otimes \sigma _{z}^{(2)},
\end{eqnarray}
with $E_{j}$$=$$E_{C}^{(j)}[1+\varsigma _{j}^{2}/(1-\varsigma
_{12}^{2})]$, and $\varsigma
_{j}$$=$$E_{J}^{(j)}/(2E_{C}^{(j)}),\,\varsigma
_{12}=E_{12}/E_{C}^{(j)}$. The evolution corresponding to this
effective Hamiltonian results in a two-qubit operation
\begin{eqnarray}
R_{z}^{(12)}(\chi)=\exp[-i\chi_{12}\sigma _{z}^{(1)}\sigma
_{z}^{(2)}]\otimes \! \! \prod_{j=1,2}\exp [-i\chi_{j}\sigma
_{z}^{(j)}],
\end{eqnarray}
where $\chi=\{\chi_{j},\chi_{12}\}$ and $\chi_j=E_{j}t/\hbar$,
$\chi_{12}=E_{12}t/\hbar$. For the simplest case where
$E_J^{(j)}=0$ and thus $\varsigma_j=0$, we have $E_j=E_{C}^{(j)}$.
By using a refocusing technique, like in NMR~\cite{GC}, we can
effectively realize another important single-qubit operation:
\begin{eqnarray}
R_{z}^{(j)}(\phi_{j})=[R_{z}^{(12)}(\chi)\sigma_{x}^{(k)}]^2=\exp
[-i\phi_{j}\sigma _{z}^{(j)}],
\end{eqnarray}
with $\phi_j=2\chi_j$. The inverse of this operation, i.e., the
gate $R_{z}^{(j)}(-\phi_{j})=\exp [i\phi_{j}\sigma _{z}^{(j)}]$
can be obtained by changing the signs of $E_j$ via controlling the
applied gate voltage.

The single-qubit gates $R_{x}^{(j)}(\varphi_j)$ and
$R_{z}^{(j)}(\phi_{j})$ do not commute, and thus constitute a
universal single-qubit gate set, which can assist the realization
of two-qubit gates to implement any unitary operation on this
circuit. For example, a Hadamard-like operation
\begin{eqnarray}
R_{j}(\theta _{j}) &=&R_{z}^{(j)}(\theta _{j}/2)R_{x}^{(j)}(\pi
/4)R_{z}^{(j)}(-\theta _{j}/2)  \notag \\
&=&\frac{1}{\sqrt{2}} \left(
\begin{array}{cc}
1 & -i \exp(i\theta _{j}) \\
-i \exp(-i\theta _{j}) & 1%
\end{array}
\right),
\end{eqnarray}
can be implemented, which will take an important role for testing
Bell's inequality.

\section{Testing Bell's inequality by using effective single-qubit local operations}

By using the above dynamical decoupling procedure, we now show
that Bell's inequality may be tested with fixed-coupling
Josephson-qubits.

First, the desired entanglement between these SQUID-based qubits
can be created in a repeatable way. Initially, the system works
sufficiently far from the co-resonance point and remains at the
state $|\psi _{0}\rangle =|00\rangle $, $\Phi_{j}=0$. Now, a pair
of gate voltage pulses brings the system to the co-resonance
point~\cite{Tsai03} and lets the system undergo the evolution
$\hat{U}_{3}(t)=\exp (-it\hat{H}_{3}/\hbar )$, with
\begin{equation}
\hat{H}_{3}=-\varepsilon _{J}\sum_{j=1,2}\sigma
_{x}^{(j)}+E_{12}\,\sigma _{z}^{(1)}\otimes \sigma _{z}^{(2)}.
\end{equation}
For simplicity, here we assume that $\varepsilon
_{J}^{(1)}=\varepsilon _{J}^{(2)}=\varepsilon _{J}$. We
analytically derive the time-dependent degree of entanglement or
\textit{concurrence}~\cite{Abouraddy01} $C_{E}(t)$ of this circuit
\begin{equation}
C_{E}(t)=\frac{1}{2}\sqrt{P^{2}(t)+Q^{2}(t)},
\end{equation}
with
\begin{eqnarray}
P(t)&=&\cos^2\vartheta(t)-\cos\varrho(t)\nonumber\\
&+&\sin^2\vartheta(t)
\left(\frac{1}{1+\tilde{\zeta}^2}-\frac{1}{1+\tilde{\zeta}^{-2}}\right),\\
Q(t)&=&\frac{\sin^2[2\vartheta(t)]}{\sqrt{1+\tilde{\zeta}^{-2}}}-\sin\varrho(t),
\end{eqnarray}
and
\begin{eqnarray*}
\vartheta(t) &=& \gamma (t)(1+\tilde{\zeta}^{2})^{1/2},\ \ \
\varrho(t)\; = \; 2\tilde{\zeta}\gamma (t), \\
\gamma(t)&=& 2\varepsilon_{J}t/\hbar,\ \ \ \
 \tilde{\zeta}=E_{12}/(2\varepsilon_{J}).
\end{eqnarray*}
%
Figure 3 shows this evolution, showing some plateaus near the
times $t_{e}$ when $\sin \vartheta (t_{e})=0$. At these times, the
system is in the following compact entangled state
\begin{equation}
|\psi _{e}\rangle =\alpha |00\rangle +\beta |11\rangle,
\end{equation}
with
\begin{equation}
C_{E}(t_{e})=2|\alpha_+ \alpha_- |=|\sin (E_{12}t_{e}/\hbar )|,
\end{equation}
and
$$
\alpha_{\pm}=[1 \pm \exp(\pm i \, t_{e} \, E_{12} \, / \, \hbar
)]/2 \; ,
$$
%
These states are very adjacent to the eigenstates of $\hat{H}_3$,
and almost do not evolve for several very short time intervals.
Thus, periodically, their degrees of entanglement are almost
unchanged, shown by the short plateaus in Fig.~3. The maximally
entangled states (corresponding to the top plateaus in Fig. 3)
occur when the pulse durations $t_{e}$ are set properly such that
the condition $\cos (E_{12}t_{e}/\hbar )=0$ is further satisfied.

Next, using dynamically-generated single-qubit operations, the
controllable variables $\{\theta_j\}$ can be encoded into the
generated entangled states, keeping the degree of entanglement
unchanged. The change of concurrence of the two-qubit entangled
state can be effectively suppressed by continuously applying
controllable single-qubit operations. This is similar to the
approaches for suppressing decoherence in open quantum systems by
using the quantum Zeno effect and the \textquotedblleft bang-bang"
decoupling method~\cite{Lidar04}. Thus, the new entangled state
\begin{equation}
|\psi _{e}^{\prime }\rangle =\prod_{j=1,2}\hat{R}_{j}(\theta
_{j})|\psi _{e}\rangle =\sum_{m,n=0,1}a_{mm}|mn\rangle,
\end{equation}
with
\begin{eqnarray*}
a_{00}&=&[\alpha-\beta \exp (i\theta _{1}+i\theta _{2})]/2,\\
a_{10}&=&[-i\alpha \exp (-i\theta _{1})-i\beta \exp (i\theta
_{2})]/2,\\
a_{01}&=&[-i\alpha \exp (-i\theta _{2})-i\beta \exp (i\theta
_{1})]/2,\\
a_{11}&=&[\alpha \exp (-i\theta _{1}-i\theta _{2})-\beta ]/2,
\end{eqnarray*}
has the same degree of entanglement as that of the
$|\psi_e\rangle$ generated above.

Finally, the correlations between the classical variables
$\{\theta_j\}$ can be measured by simultaneously detecting the
populations of qubits in the excited $|1\rangle$ or ground states
$|0\rangle$~\cite{martinis05}. Experimentally, the above steps can
be repeated many times for the same
classical variables and then the correlation function $E_e(\theta_1,%
\theta_2) $ can be measured as
\begin{equation}
E_e(\theta_1,\theta_2)=\frac{N_{\mathrm{same}}(\theta_1,\theta_2)-N_{\mathrm{%
diff}}(\theta_1,\theta_2)}{N_{\mathrm{same}}(\theta_1,\theta_2)+N_{\mathrm{%
diff}}(\theta_1,\theta_2)},
\end{equation}
with $N_{\mathrm{same}}(\theta_1,\theta_2)$ ($N_{\mathrm{diff}%
}(\theta_1,\theta_2)$) being the number of events with two qubits
found in the same (different) logic states. Theoretically, the
above projected measurements can be expressed via
\begin{eqnarray}
\hat{P}_T&=&|11\rangle\langle 11|+|00\rangle\langle
00|-|10\rangle\langle 10|-|01\rangle\langle
01|\nonumber\\&=&\hat{\sigma}_{z}^{(1)}\otimes
\hat{\sigma}_{z}^{(2)},
\end{eqnarray}
and the correlation in the outcomes can be calculated as
\begin{eqnarray}
E(\theta_1,\theta_2)&=&\langle\psi^{\prime}_e|\hat{P}_T|\psi^{\prime}_e\rangle\nonumber\\
&=&\pm\sin(t_eE_{12}/\hbar)\sin(\theta _1+\theta_2).
\end{eqnarray}

For the sets of angles: $\{\theta_{j},\theta^{\prime}_j\}=\{-\pi
/8,\,3\pi /8\}$, the Clauser, Horne, Shimony and Holt
(CHSH)~\cite{asp82} function
\begin{eqnarray}
f(|\psi^{\prime}_e\rangle)&=&\left|E(\theta _{1},\theta
_{2})+E(\theta _{1}^{\prime },\theta _{2})+E(\theta _{1},\theta
_{2}^{\prime
})-E(\theta _{1}^{\prime },\theta _{2}^{\prime })\right|\nonumber\\
&=&2\sqrt{2}\,\left|\sin(t_eE_{12}/\hbar)\right|
\end{eqnarray}
is larger than $2$ for
\begin{equation}
\left|\sin(t_eE_{12}/\hbar)\right|>\frac{1}{\sqrt{2}}.
\end{equation}
Therefore, properly setting the pulse duration $t_e$ to prepare
the desired entangled state (whose plateau-like concurrence is
larger than $0.707$), the CHSH-type Bell's inequality~\cite{asp82}
\begin{equation}
f_e(|\psi^{\prime}_e\rangle)<2
\end{equation}
can be effectively tested by experimentally measuring the CHSH function: $%
f_e(|\psi^{\prime}_e\rangle)=|E_e(\theta_1,\theta_2)+E_e(\theta^{\prime}_1,%
\theta_2)+E_e(\theta_1,\theta_2^{\prime})
-E_e(\theta_1^{\prime},\theta_2^{\prime})|$.
\begin{figure}[tbp]
\vspace{-0.2cm}
\includegraphics[width=13.5cm]{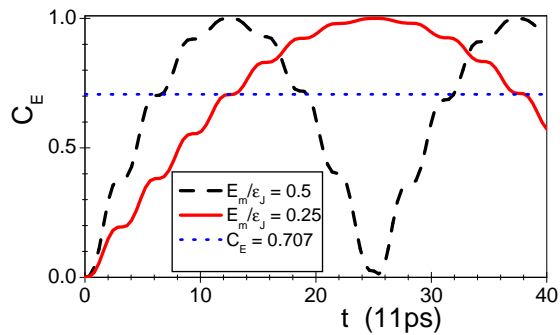}
\vspace{-5cm} \caption{Dynamical evolution of the degree of
entanglement $C_{E}(t)$ for the circuit (Fig.~1) with couplings:
$E_{m}/\protect\varepsilon _{J}=1/2$ (dashed line),\,$1/4$ (solid
line), respectively. Here, $\protect\varepsilon
_{J}=30\,\protect\mu \mathrm{eV}$. The $C_{E}=0.707$ (dotted line)
gives the threshold for the violation of Bell's inequality for the
entangled state (8), whose degree of entanglement slowly changes
during several short time intervals (see the plateaus in the
figure).}
\end{figure}

\section{Discussions and Conclusions}

The simplest dynamical-decoupling process proposed above consists
of two $\sigma_x^{(j)}$-pulses and two delays
\begin{equation}
\hat{U}_d(\tau)=\exp[-iE_{12}\tau\sigma_z^{(1)}\sigma_z^{(2)}].
\end{equation}
The duration of the $\sigma_x^{(j)}$-pulse is
calculated~\cite{Tsai03} as $t_x$$=$$(2l+1)t_0,\,l$$=$$0,1,2,...;$
and $t_0$$=$$\pi\hbar/[2\varepsilon_J^{(j)}(1+2\zeta^2_j)]\simeq
31\,\mathrm{ps}$. Consequently, the longest time delay $\tau$
between the two $\sigma_x$-pulses could be estimated~\cite{Tsai03}
as $\tau\,\sim 270\,\mathrm{ps}$ (for\, $l$$=$$0$), $\sim
200\,\mathrm{ps}$ (for\, $l$$=$$1$), $\sim 145\,\mathrm{ps}$
(for\, $l$$=$$2$), and $\sim 98\,\mathrm{ps}$ (for\, $l$$=$$3$),
etc. Thus, it is easy to experimentally check this simplest
proposal for eliminating the fixed interbit coupling by using the
pulse sequence:
$\hat{U}_d(\tau)\sigma_x^{(j)}\hat{U}_d(\tau)\sigma_x^{(j)}$, or
$\sigma_x^{(j)}\hat{U}_d(\tau)\sigma_x^{(j)}\hat{U}_d(\tau)$.
After these operations the two qubits should return to their
initial states. Furthermore, an universal two-qubit
controlled-$\sigma_z^{(j)}$ gate could be implemented by using the
operational sequence:
$\hat{U}_d(-\pi/4E_{12})R_z^{(k)}(\pi/4)R_z^{(j)}(\pi/4)$.

Similar to other theoretical schemes (see,
e.g.,~\cite{makhlin99}), the realizability of the present proposal
also faces the technological challenge of rapidly on/off switching
the Josephson energy of the qubit by using fast magnetic
pulses~\cite{Uwazumi02}. This experimental difficulty could be
relaxed by increasing the durations of the applied pulses.
Especially, the decoherence time of the two-qubit
capacitively-coupled Josephson circuit~\cite{Makus03} could be
increased by decreasing the coupling capacitance $C_{m}$. In
principle, the lifetime of the generated entangled state (19)
adequately allows to perform the required operations for testing
Bell's inequality (27), since such a state is very adjacent to the
eigenstates of the circuit's Hamiltonian $\hat{H}_3$. In fact, the
decay time of a two-qubit excited state is long (up to $\sim 0.6$
ns), even for very strong interbit coupling (e.g., $E_m\simeq
2\varepsilon_J^{(j)}$ in the recent experiment~\cite{Tsai03}). In
addition, the influence of the environmental noise and operational
imperfections is not fatal, as the nonlocal correlation
$E(\theta_i,\theta_j)$ in Bell's equality is {\it
statistical}---i.e., its fluctuations could be effectively
suppressed by averaging over several repeatable experiments.

In summary, we propose an effective dynamical-decoupling approach
to overcome the fixed-interaction difficulty in superconducting
nanocircuits. The dynamically-generated single-qubit operations
may be used to test Bell's inequality, providing another way to
verify the existence of entanglement between two capacitively
coupled Josephson qubits. The proposed approach can be easily
modified to manipulate quantum entanglement in other
"fixed-interaction" solid-state systems, e.g., the capacitively
(inductively) coupled Josephson phase (flux) circuits, and the
Ising (Heisenberg)-spin chain.

\section*{Acknowledgments}

We thank Drs. J. S. Tsai, Y. Pashkin, and X. Hu for useful
discussions. This work was supported in part by the National
Security Agency (NSA) and Advanced Research and Development
Activity (ARDA) under Air Force Office of Research (AFOSR)
contract number F49620-02-1-0334, and by the National Science
Foundation grant No. EIA-0130383.

\end{document}